\magnification=1200 \baselineskip=28 truept plus1pt minus1pt
\parskip=6 true pt
 \hoffset=0.2 true in
\hsize=6.2 true in  \nopagenumbers
\def\section#1{\goodbreak\medbreak\noindent{\tenbf #1}\nobreak\smallskip}
\def\subsection#1{\goodbreak\smallbreak\noindent{\bf#1}\nobreak\smallskip}
\def\ref#1#2#3#4#5#6{\medbreak\noindent\item{#1}{#2},\ {#3}\
{\it #4}{#5}{#6}}

 \font\bigbold=cmbx10 scaled\magstep1

\def\romone{\uppercase\expandafter{\romannumeral 1}}
\def\romtwo{\uppercase\expandafter{\romannumeral 2}}
\def\romtre{\uppercase\expandafter{\romannumeral 3}}
\def\romfor{\uppercase\expandafter{\romannumeral 4}}
\def\romfiv{\uppercase\expandafter{\romannumeral 5}}
\def\romsix{\uppercase\expandafter{\romannumeral 6}}
\def\romsev{\uppercase\expandafter{\romannumeral 7}}
\def\romeig{\uppercase\expandafter{\romannumeral 8}}
\def\romnin{\uppercase\expandafter{\romannumeral 9}}

\def\caption #1! #2!{\noindent{\bf #1\enspace}{\sl #2}}

\def\toclin1#1#2{\line{\noindent #1\ \dotfill\ #2}}
\def\toclin2#1#2{\line{\noindent #1\ \dotfill\ #2}}

\input epsf

\centerline{\bigbold  Focusing, Power Tunneling and Rejection from }
\centerline{\bigbold Chiral and/or Chiral Nihility/Nihility
Metamaterials Layers} \centerline{\bigbold By } \vskip .05 true in
 \centerline{ Syed Touseef Hussain Shah$^1$, Faiz Ahmad$^2$, Aqeel A.
 Syed$^3$, Qaisar Abbas Naqvi$^4$
} \centerline{ ${}^1$touseefshah1 @ yahoo.com, ${}^2$faizsolangi @
gmail.com,}
 \centerline{ ${}^3$aqeel @ qau.edu.pk, ${}^4$qaisar @ qau.edu.pk}
\vskip .05 in {\baselineskip 14pt\centerline{${}^{1,3,4}$ Department
of Electronics, Quaid-i-Azam University, 45320, Islamabad} }
\centerline{${}^2$ Department of Electronics, CIIT, Islamabad}

\centerline {\bf ABSTRACT}

Focusing of electromagnetic plane wave  from  a large size
paraboloidal reflector, composed of layers of chiral and/or chiral
nihility metamaterials, has been studied using Maslov's method.
First, the transmission and reflection of electromagnetic plane wave
from two parallel  layers of chiral and/or chiral nihility
metamaterials are calculated using transfer matrix method. The
effects of change of angle of incidence, chirality parameters and
impedances of layers are noted and discussed. Special cases by
taking very large and small values of permittivity of second layer,
while assuming value of corresponding chirality equal to zero, are
also treated. These special cases are equivalent to reflection from
a perfect electric conductor backed chiral layer and nihility backed
chiral layer, respectively. Results of reflection from parallel
layers have been utilized to study focusing from a large size
paraboloidal reflector. The present study, on focusing from a
paraboloidal reflector, not only unifies several published works
conducted by different researchers but also provides better
understanding of new cases.

 \vfill\eject

\section{1.0 Introduction }
Chiral  media, have been well  known  for  a  long  time   due to
some
             interesting properties, exhibiting in  the
             optical  range of frequencies. Out of these properties, optical
             activity and circular dichroism are of practical interest~[1-6].
             Chiral medium can be thought as  composed of
             numerous randomly  oriented  chiral
             objects which  can  never  be brought into  congruence  with  their  mirror
             images  by  any translation  or  rotation~[7-10].  Many chiral objects are
             found in nature such as irregular tetrahedrons, sugar molecules and wire
             helices.
             In fact the word ``chirality" is derived from the greek word ``cheir" having meaning of
             hand; an object possessing above mentioned property.
             In science particular word ``chirality" instead of
             dissymmetry was first introduced by Lord Kelvin, Professor of natural philosophy in the University of
             Glasgow. Now this word is used to describe
             the media microscopically composed of the chiral objects.

              When linearly polarized wave falls on a slab of chiral
             medium, it splits into two circularly polarized waves:
             one left circularly polarized and the other
             right circularly polarized~[8-10]. After passing through the slab the
             two  waves combine to yield a linearly polarized wave
             whose plane of polarization is rotated with respect to
             the plane of polarization of the incident wave. Effect of chirality of the medium which rotates the plane of
             linearly polarized wave passing through it is studied
             by Biot, Arago and Friesnel. This phenomenon is named as optical activity. The amount of rotation depends
              upon the distance traveled by the wave in the slab. This phenomenon is named as circular birefringence.
              Circular dichroism represents the phenomenon of
              different amount of field absorbtion for the
              left and right circularly polarized waves as they pass through the chiral media.
              The constitutive relations for chiral
             metamaterial~[8] are given below,
             $$\eqalignno{{\bf D} &=\epsilon {\bf E}+i{\kappa}{\bf H} &(1)\cr
             {\bf B} &=\mu{\bf H}-i{\kappa}{\bf E} &(2)}$$
              The behavior of propagation and radiation of the field
               in a chiral medium has been investigated
              by several authors ~[8-10].

              Left-handed materials are those materials in which electric, magnetic
              and the wave vector follow the left hand rule. Vaselago introduced
              the concept of negative refraction  for the materials
              obeying left hand rule~[11]. In chiral medium, negative real parts
              of the permittivity
              and permeability lead an isotropic chiral medium to exhibit circular dichroism that is
              reverse with respect to that exhibited by an identical medium but with positive real parts of permittivity
              and permeability. These left handed materials exhibit some interesting properties such
              as  backward wave and double negative parameters~[12]. Negative refraction can also
              be achieved by using the chiral metamaterials~[13-16].  In chiral matematerials strong
              enough chirality produces negative refractive
              index for one of the eigenwaves~[17-18].

 In electromagnetics, Lakhtakia introduced the concept of nihility metamaterial for such a material whose
  permittivity and permeability approaches to zero~[19].
 Constitutive relations for nihility metamaterial are,
$$\eqalignno{{\bf D}&=0 &(3)\cr
{\bf B}&=0 &(4)}$$ Lakhtakia showed that propagation is not possible
in nihility metamaterial.
 Later, Tretyakov et al. [20] extended this concept
 of nihility for the isotropic chiral medium. A special case of chiral metamaterial is called chiral nihility
 metamaterial for which at certain value of frequency, real parts of permittivity
and permeability  both are simultaneously zero. That is,
$\epsilon\rightarrow0$, $\mu\rightarrow0$ and $\kappa\neq 0$ at
nihility frequency.   Constitutive relations for chiral nihility
metamaterials become as,
$$\eqalignno{{\bf D} &=i{\kappa}{\bf H} &(5)\cr
{\bf B} &=-i{\kappa}{\bf E} &(6)}$$ Chiral nihility metamaterial has
two wavenumbers of equal magnitude and opposite signs. When a
dielectric-chiral nihility half space is excited by an oblique
incident linearly polarized plane wave, two circularly polarized
plane waves having opposite handedness are produced. One of them is
forward wave while the other is backward wave which yields
phenomenon of negative refraction in chiral nihility metamaterial.

If a perfect electric conductor (PEC) interface placed in chiral
nihility metamaterial is excited by a forward plane wave. Only one
reflected wave parallel to the incident wave is produced, i.e.,
negative reflection happens. The reflected wave is backward wave
which cancels the incident forward wave to produce zero power
propagation. Interestingly, if a chiral nihility slab backed by PEC
interface is excited by linearly polarized plane wave, effect of
each eigenwave in chiral nihility after reflection from PEC
interface is canceled leading to a situation as if front face of the
geometry is PEC and there is no chiral slab~[21]. A PEC waveguide
coated with chiral nihility metamaterial confines propagation of
power within un-coated region of the guide~[22].

In electromagnetics, Brewster angle is defined as angle of incidence
for which reflection power is zero. In all introductory text books
on electromagnetic theory ~[23,24], reflection of vertically
polarized (parallel polarized  or transverse magnetic~(TM)) plane
wave and horizontally polarized (perpendicular polarized or
transverse electric~(TE)) plane wave from a dielectric-dielectric
half space boundary is usually presented to explain the concept of
Brewster angle. It is mentioned that vertically polarized wave
experiences zero reflection at one incident angle whereas for all
angles, reflection is non-zero for horizontal polarization. This
means that if dielectric-dielectric half space is excited by a wave,
having both the vertical and horizontal polarizations, at the
Brewster angle the reflected wave will be linearly polarized with
horizontal polarization only. The angle of incidence that allows
total reflection of power from a planar dielectric-dielectric
interface is known as critical angle. Critical angle exists only for
perpendicular polarization if the wave propagates from a  denser
dielectric medium to a rare dielectric.

In the same context, reflection of plane wave from an interface of
an achiral dielectric and a chiral/chiral nihility metamaterial was
discussed by Qiu et al.~[25]. Their study reveals that for certain
values of constitutive parameters, the results similar and opposite
to that of conventional dielectric-dielectric interface can also be
achieved, that is, existence of zero reflection for only parallel
polarization in the dielectric-chiral case. They also showed that,
for certain values it is possible to have no-Brewster angle for both
polarizations and total reflection can be achieved for a wide range
of incident angles. Power corresponding to the electromagnetic waves
associated with a planar interface of two chiral and/or chiral
nihility metamaterials was studied by Faiz et al.~[26] when it is
excited by a plane wave. Using numerical results, interesting
characteristics such as complete power transmission/rejeection and
band pass/band reject filter were observed.

\section{1.1. GO Field and Maslov's Method}

Geometrical optics~(GO) is powerful tool for the study of wave
motion at high
 frequencies~[27-29], however it fails at the
the caustics. A caustic is a region where area of a ray tube is zero
and hence introduces a mathematical singularity when ever intensity
per unit area is needed to be counted though physically this is not
the case. Physically the field is always finite at caustics and
caustics are of practical interest in many applications including
defense and medical sciences.
 Maslov proposed an alternative
 method to find the fields in the caustic
  region~[30]. Maslov's method combines the simplicity of asymptotic ray theory and
the generality of the Fourier transform method. This is achieved by
representing the GO fields in terms of mixed coordinates consisting
of wave vector coordinates and space coordinates. Maslov's method
are applied to study the field near the caustics of the focusing
systems by many authors~[31-40]. Focusing from a PEC cylindrical and
spherical reflectors using Maslov's method was first treated by
Hongo and Ji~[33]. Faryad and Naqvi extended this work for reflector
coated with chiral metamaterial~[39]. Illahi and Naqvi studied the
focusing from chiral nihility coated PEC and perfect electromagnetic
conductor (PEMC) cylindrical reflector~[41]. In present work, one
interest is also to unify all these cases so that these become
special cases of the geometry, under investigation.

Consider three dimensional vector wave equation, in cartesian
coordinates $r=(x,y,z)$, describing the field in medium having
wavenumber $k_0$
$$ \nabla^2 {\bf U}(r) + k_0^2 {\bf U}(r) =0$$
Expressing the solution ${\bf U}(r)$ of wave equation in terms of
well known Luneberg-Kline series yields Eikonal equation for phase
$s(r)$ and transport equation for amplitude. For homogeneous and
lossless medium, Eikonal equation reduces to Hamiltonian equation as
$$H({\bf {r,p}})=({\bf {p}.\bf{p}}-1)/2=0, \qquad\qquad {\bf{p} }=\nabla s $$
where ${\bf p}$ is the wave vector.
 The solution of Hamiltonian equation  is given below
$$x=\xi+p_{x}t,\quad y=\eta+p_{y}t, \quad
z=\zeta+p_{z}t,\qquad
 p_{x}=p_{x_{0}},\quad p_{y}=p_{y_{0}},\quad
p_{z}=p_{z_{0}} $$ where $(\xi,\eta,\zeta)$ and $(p_{x_{0}},
p_{y_{0}}, p_{z_{0}})$ are the initial values of cartesian
coordinates $(x,y,z)$ and wave vector coordinates $(p_{x}, p_{y},
p_{z})$ respectively and $t$ is parameter along the ray. Finite
field around caustic may be obtained using following expression
obtained from Maslov's method~[34,40] $$ \eqalignno {{\bf
U}{(r)}&={k_{0}\over2\pi} \int_{-\infty}^\infty \!
\int_{-\infty}^\infty \! {\bf E}(r_0)\left[{{D(t)\over D(0)}}
{\partial(p_{x},p_{y})\over
\partial(x,y)}\right]^{-1/2}&\cr
&\exp(-i k_{0}(s_{0}+t  -x(p_{x}, p_{y},z)p_{x}- y(p_{x},p_{y},
z)p_{y}+xp_{x}+yp_{y}))dp_{x}dp_{y} & }$$ $s_0$ is the initial
phase. Above equation provides uniform solution around the caustic
region.

In this paper, transmitted and reflected  powers from two parallel
layers of chiral and/or chiral nihility metamaterials with respect
to angle of incidence and chiralities of the layers are analyzed.
Limiting cases of permittivity, with value of chirality equal to
zero, are also taken into account and their interpretation are
given. Focusing from a large size paraboloidal reflector, composed
of two layers of chiral and/or chiral nihility metamaterials, is
studied by utilizing the Maslov's method to give the remedy of GO
which fails at caustic.

\section {2.0. Formulation}
Consider a geometry consisting of two infinite parallel
layers of chiral metamaterials placed in air as shown in Fig.~2.1.
Physical width of the two layers is taken as $d_{1}$ and $d_{2}$,
respectively. Constitutive parameters for the two layers
 are denoted by $(\epsilon_{1}, \mu_{1}, \kappa_{1})$ and
$(\epsilon_{2}, \mu_{2}, \kappa_{2}),$ respectively.  Air in which
layers are placed have constitutive parameters $\epsilon_{0}$ and
$\mu_{0}$.

The above mentioned geometry is excited by an oblique incident
linearly polarized electromagnetic plane wave. $\theta_{i},$
$\theta_{r}$ and $\theta_{t}$ are the angle of incidence, reflection
and transmission with respect to the walls of layers, respectively.
As chiral medium supports two circularly polarized eigen waves
having different wavenumbers so fields in each layer are written as
linear combination of left circularly polarized (LCP) and right
circularly polarized (RCP) plane waves. Both LCP and RCP waves
propagate in forward as well as backward direction in each layer.
The wavenumbers of both eigen waves in a chiral media are different.
Refractive indices, wavenumbers and intrinsic impedance of the
medium for first chiral layer are,
$$\eqalignno{ n_{1(R,L)}
&=\sqrt{\epsilon_{r1}\mu_{r1}}\pm\kappa_{1} &(7)\cr
k_{1(R,L)}&=\omega\left({\sqrt{\epsilon_{1}\mu_{1}}\pm\kappa_{1}}\right)
&(8)\cr \eta_1&=\sqrt{\mu_1\over \epsilon_1} &(9)}$$ where $n_{1R}$
and $k_{1R}$ are refractive index and wavenumber for RCP wave,
respectively. Quantity $\epsilon_{r1}=\epsilon_1/\epsilon_0$ is the
relative permittivity of the medium occupying first layer. \vskip
-.04 in \centerline {\epsfxsize 3 in \epsfysize 3.0
in\epsfbox{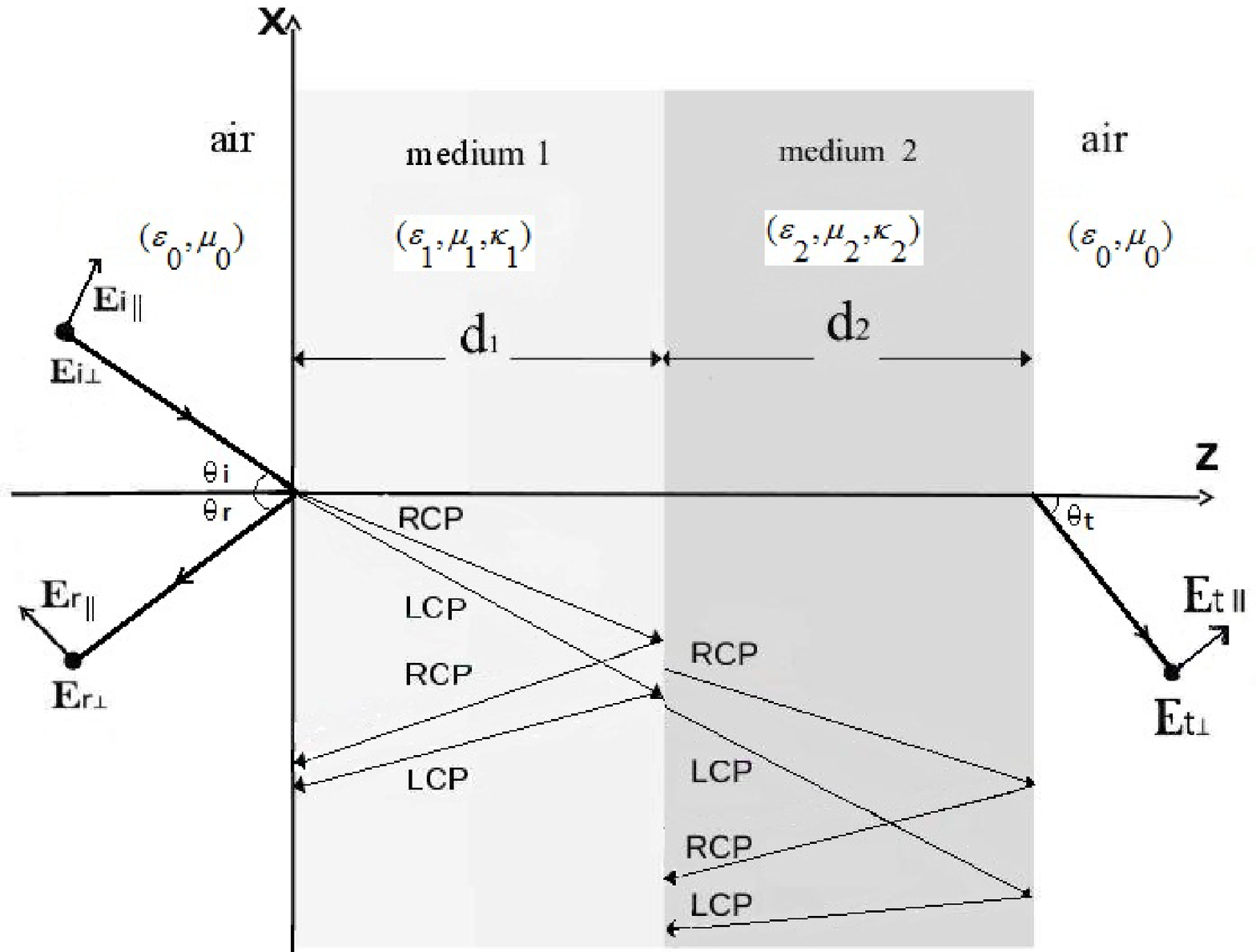}} \vskip -.2 in \caption Figure 2.1.! {Fields
representation inside and outside the layers.}!

For second chiral layer, these quantities are,
$$\eqalignno{n_{2(R,L)}&=\sqrt{\epsilon_{r2}\mu_{r2}}\pm\kappa_{2} &(10)\cr
k_{2(R,L)}&=\omega\left({\sqrt{\epsilon_{2}\mu_{2}}\pm\kappa_{2}}\right)&(11)\cr
\eta_2&=\sqrt{\mu_2\over \epsilon_2} &(12)}$$ Wavenumber and
intrinsic  impedance of the host medium are given below,
$$\eqalignno{  \eta_0&=\sqrt{ \mu_0/\epsilon_0}&(13)\cr
k_0&=\omega\ \sqrt{ \mu_0\epsilon_0}&(14)}$$

 Expressions for the incident and reflected fields and
the field transmitted after passing through both the layers are,
$$\eqalignno{ {\bf E}_{\rm inc}&=[  E_{i\|}({\bf a}_z \sin \theta_{i}+{\bf
a}_x\cos\theta_i)+E_{i\bot}{\bf a}_y] \exp[-ik_0(z \cos\theta_i-x
\sin\theta_i)] &(15)\cr \bf {H}_{\rm inc}&=1/\eta_0[-E_{i\bot}({\bf
a}_z\sin\theta_i+{\bf a}_x\cos\theta_i)+E_{i\|}{\bf a }_y]
\exp[-ik_0(z \cos\theta_i-x \sin\theta_i)]
  &(16)\cr \bf {E}_{\rm ref}&=[ E_{r\|}(-{\bf a}_z\sin\theta_r+{\bf
a}_x\cos\theta_r)+E_{r\bot}{\bf a}_y] \exp[ik_0(z \cos\theta_r+x
\sin\theta_r)]&(17)\cr
  \bf {H}_{\rm ref}&=1/\eta_0[E_{r\bot}(-{\bf a}_z\sin\theta_r+{\bf
a}_x\cos\theta_r)-E_{r\|}{\bf a}_y] \exp[ik_0(z \cos\theta_r+x
\sin\theta_r)] &(18)\cr
  \bf{E}_{\rm tra}&=[E_{t\|}({\bf a}_z\sin\theta_t+{\bf
a}_x\cos\theta_t)+E_{t\bot}{\bf a}_y] \exp[-ik_0(z \cos\theta_t-x
\sin\theta_t)]&(19)\cr  \bf{H}_{\rm tra}&=-1/\eta_0[ E_{t\bot}({\bf
a}_z\sin\theta_t+{\bf a}_x\cos\theta_t)-E_{t\|}{\bf a}_y]
\exp[-ik_0(z \cos\theta_t-x \sin\theta_t)]&(20)}$$
  In above equations, $E_{r\|}$,
$E_{r\bot}$, $E_{t\|}$, and $E_{t\bot}$ are unknown coefficients
corresponding to parallel and perpendicular components of the
respective fields.

Expressions of fields inside each chiral layer can be written as,
$$\eqalignno{ \bf { E}_{p+}&= E_{fL}[({\bf a}_x \cos\theta_{pL}+ {\bf
a}_z\sin\theta_{pL})+i{\bf a}_{y}] \exp[-ik_{pL}(z \cos\theta_{pL}-x
\sin\theta_{pL})] \cr & +  E_{fR}[({\bf a}_x\cos\theta_{pR}+{\bf
a}_z\sin\theta_{pR})- {i\bf a}_y] \exp[-ik_{pR}(z \cos\theta_{pR}-x
\sin\theta_{pR})] &(21)\cr \bf {H}_{p+}&=-( i/{\eta_ p})
E_{fL}[({\bf a}_x\cos\theta_{pL}+{\bf a}_z\sin\theta_{pL})+i{\bf
a}_y]\cr &\times\exp[-ik_{pL}(z \cos\theta_{pL}-x \sin\theta_{pL})]
+( i/{\eta_ p}) E_{fR}[({\bf a}_x\cos\theta_{pR}+ {\bf
a}_z\sin\theta_{pR})-i{\bf a}_y]\cr &\times \exp[-ik_{pR}(z
\cos\theta_{pR}]-x \sin[\theta_{pR})] &(22)\cr \bf{E}_{p-}&=
E_{bL}[(-{\bf a}_z\sin\theta_{pL}+{\bf a}_x\cos\theta_{pL})+i{\bf
a}_y] \exp[ik_{pL}(z \cos\theta_{pL}+x \sin\theta_{pL})] \cr &+
E_{bR}[(-{\bf a}_z\sin\theta_{pR}+{\bf a}_x\cos\theta_{pR})-i{\bf
a}_y] \exp[ik_{pR}(z \cos\theta_{pR}+x \sin\theta_{pR})] &(23) \cr
\bf {H}_{p-}&=-( i/{\eta_ {p}}) E_{bL}[(-{\bf
a}_z\sin\theta_{pL}+{\bf a}_x\cos\theta_{pL})+i{\bf a}_y]\cr&\times
\exp[ik_{pL}(z \cos\theta_{pL}+x \sin\theta_{pL})]+( i/{\eta_ p})
E_{bR}[(-{\bf a}_z\sin\theta_{pR}+ {\bf a}_x\cos\theta_{pR})-i{\bf
a}_y]\cr &\times \exp[ik_{pR}(z \cos\theta_{pR}+x
\sin\theta_{pR})]&(24)}$$ where $ k_{pL}$ and $ k_{pR}$ represent
the wavenumbers for LCP and RCP waves, respectively. Value of $p$,
as 1 and 2, describes fields for  first layer and the second layer,
respectively. Subscripts $b$ and $f$ are used to represent forward
and backward waves, respectively. In above equations, $E_{fL}$,
$E_{fR}$, $E_{bL}$, and $E_{bR}$ are unknown coefficients.
Tangential components of fields are continuous at interfaces located
at $z=0, d_1, d_1+d_2$ and these boundary conditions are used to
find the unknown coefficients. Snell's law is used to determine the
relation among the angles of incidence, reflection and transmission
by using following relations,
$$\eqalignno{ \theta_{r}&=\theta_{i}\cr k_0
\sin\theta_i=k_{1(R,L)}\sin{\theta_{1(R,L)}}&=k_{2(R,L)}\sin{\theta_{2(R,L)}}=k_{t}
\sin\theta_t}$$ where $k_{0}$ and $k_{t}$ denote the wavenumbers for
medium before and after the layers and are equal when these two
layers are placed in air.  Using transfer matrix method (TMM)
discussed in [42], three matching matrices each relating the fields
at an interface are obtained. Two 4$\times$4 propagation matrices,
linking fields between two consecutive interfaces are obtained.
Numerical results for powers corresponding to parallel and
perpendicular components of transmitted and reflected fields are
plotted. In section~2, it is assumed that optical width of each
layer is $\lambda_{0}/4$, where $\lambda_{0}$ is wave length in air
corresponding to the operating frequency.
\section{2.1. Results and Discussion}
In this section, numerical results for powers corresponding to
parallel and perpendicular components of
 the reflected and transmitted fields are presented and analyzed. Four different cases, summarized below, are considered for this purpose.
  \item{i.} Both layers are of chiral metamaterials  (c-c)
  \item{ii.} First layer is of chiral nihility and second is of chiral metamaterials (cn-c)
  \item{iii.} First layer is of chiral  whereas  second is of chiral nihility metamaterials (c-cn)
  \item{iv.} Both layers are of chiral nihility metamaterials (cn-cn).

\noindent Results corresponding to above cases are
  shown by Figure 2.2 to Figure~2.9.

  In addition to the above, the limiting cases of permittivity of
  second layer keeping $\kappa_2=0$ are also considered
  for the discussion. These cases are,
  \item{a.} First layer is of chiral metamaterial whereas permittivity of second layer is taken very large with chirality $\kappa_2$ equal to zero (c-PEC)
  \item{b.} First layer is of chiral nihility metamaterial whereas permittivity of second layer is taken very large with chirality $\kappa_2$ equal to zero (cn-PEC)
  \item{c.} First layer is of chiral metamaterial whereas permittivity of second layer is taken very small with chirality $\kappa_2$ equal to zero (c-n)
   \item{d.} First layer is of chiral nihility metamaterial whereas permittivity of second layer is taken very small with chirality $\kappa_2$ equal to zero (cn-n)

\noindent Results corresponding to above situations are
  shown by Figure 2.10 to Figure~2.13.
$\|$ and $\bot$ signs are used to represent power for parallel and
perpendicular components, respectively. For example: ${\rm
c-c}_{\|}$ and ${\rm c-c}_{\bot}$ represent  parallel and
perpendicular components of power for chiral-chiral cases,
respectively.

In Figure~2.2, powers corresponding to parallel and perpendicular
components of reflected fields versus angle of incidence are shown.
It is assumed that the intrinsic impedances of metamaterials filling
both layers are same.
 Figure contains plots for four different cases, i.e.,
c-c, cn-c, c-cn, and cn-cn. Brewster angle is defined as the angle
of incidence for which reflection of power, both parallel and
perpendicular components, is zero. Reflected powers for c-c case are
zero for $\theta_i<23^0$ and this gives a range of brewster angles.
In the case of cn-cn, a range of Brewster angles exists
 for $\theta_i<17.5^0$. After this range, power for parallel component
increases whereas power for perpendicular component remains
negligible for the entire range of incident angles. In the case of
cn-c, parallel component is nonzero everywhere and near grazing
incidence almost total reflection of power happens whereas power for
perpendicular component remains negligible through the entire range
of incident angles. It may be noted that range of Brewster angles
exist only for c-c and cn-cn cases and range of angles yielding
almost total reflection exists only for parallel component of cn-c
case. For c-cn case, power only for perpendicular component is zero
for $\theta_i<27.5^0$ and corresponding parallel component is
nonzero everywhere.

Figure~2.3 describes behavior of the parallel and perpendicular
components of transmitted powers for impedance matching of both
layers. For c-c case, power for parallel and perpendicular
components have same initial amplitude, i.e., 0.5. Considering cn-c
case, power of both parallel and perpendicular components of
transmitted power follow each other.  In c-cn case, perpendicular
component is dominant over parallel component whereas for cn-cn
case, it is observed that power of parallel component  is dominant
over the perpendicular component.

\bigskip
\centerline {\epsfxsize 3.5 in \epsfysize 2.5 in\epsfbox{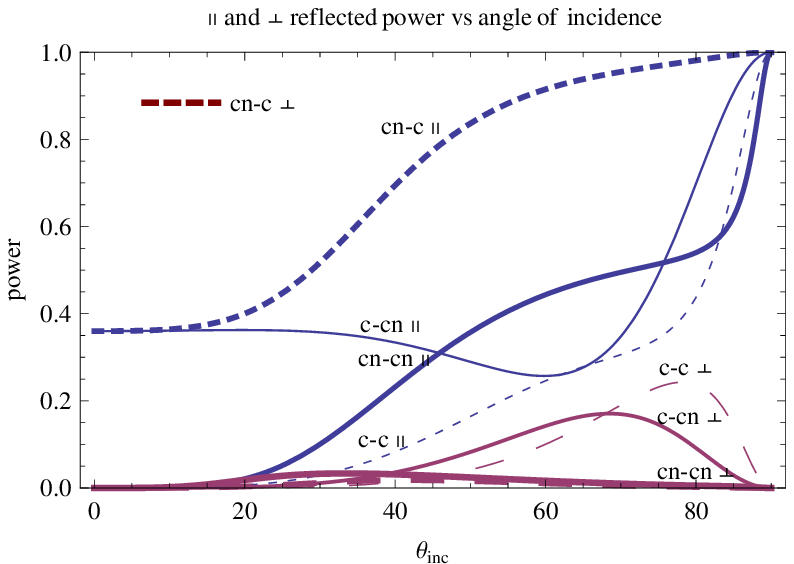}}
 \vskip -0.1 in
\caption Figure 2.2.! {Behavior of the reflected power for parallel
and perpendicular field components. $\kappa_{1}=.25,$
$\kappa_{2}=.75,$ $\eta_{1}$=$\eta_{2}=2$ }!

Figure~2.4 shows the behavior of reflected power for layers having
different intrinsic impedances. For c-c and cn-c cases, no Brewster
angle exists and power for parallel component is dominant over
corresponding perpendicular component. For c-c case, reflected power
only for perpendicular component is zero for $\theta_i<34.3^0$ and
$\theta_i>85^0$. For cn-c case, when $\theta_i>22^0$, power for
parallel component of reflected field starts increasing and after
$\theta_i=80^0$, almost total reflection is observed because
perpendicular component has insignificant reflected power.
Perpendicular component has very low value between
$\theta_{i}$=$34.3^0$ and $\theta_{i}$=$78^0$ and is zero elsewhere.
Both c-cn and cn-cn cases have same range of Brewster angles, i.e.,
$\theta_{i}\leq$ $11.46^0$.

Figure~2.5 is about   transmitted power in case of impedance
mismatch. For c-c case, both components of power have equal
contribution at $\theta_i=31.5^0$.  For cn-c and cn-cn cases, power
for parallel component has higher values and follows the power for
corresponding perpendicular component.

\vfill\eject
\bigskip
\centerline {\epsfxsize 3.5 in \epsfysize 2.5 in\epsfbox{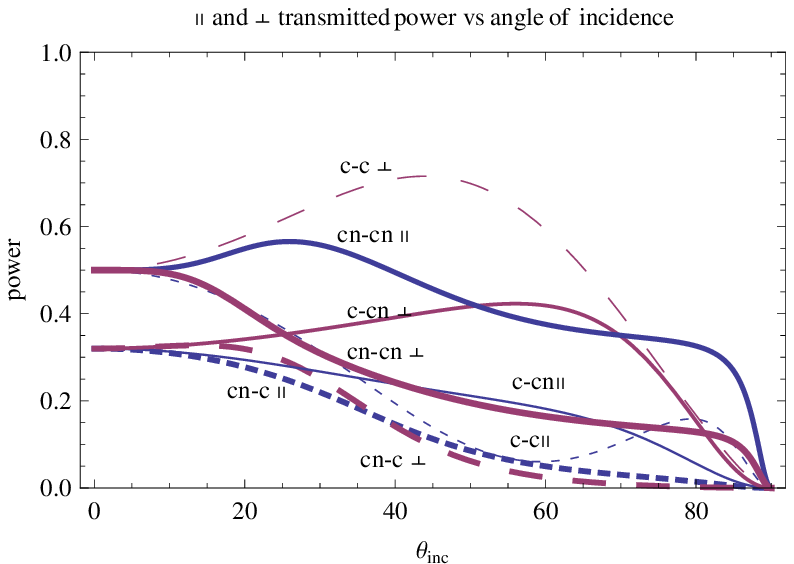}}
 \vskip -0.1 in
\caption Figure 2.3.! {Behavior of transmitted power for parallel
and perpendicular field components. $\kappa_{1}=.25$
$\kappa_{2}=.75,$ $\eta_{1} $=$\eta_{2}=2$ }!

\bigskip
\bigskip
\bigskip
\bigskip

\centerline {\epsfxsize 3.5 in \epsfysize 2.5 in\epsfbox{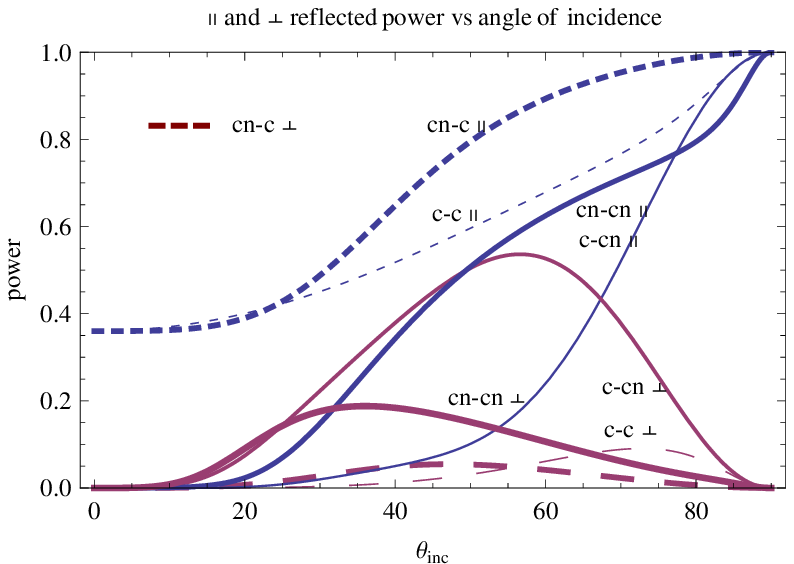}}
 \vskip -0.1 in
\caption Figure 2.4.! {Behavior of reflected power for parallel and
perpendicular field components. $\kappa_{1}$= $\kappa_{2}=.25,$
$\eta_{1}=1 $  and   $\eta_{2}=2$ }!

 \vfill\eject
\bigskip
\centerline {\epsfxsize 3.5 in \epsfysize 2.5 in\epsfbox{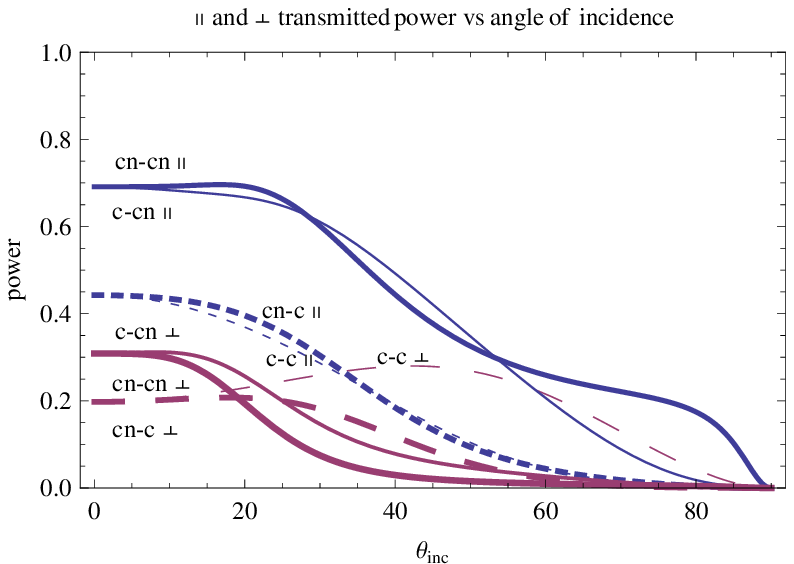}}
 \vskip -0.1 in
\caption Figure 2.5.! {Behavior of transmitted power for parallel
and perpendicular field components. $\kappa_{1}$=$\kappa_{2}=.25,$
$\eta_{1}=1 $  and   $\eta_{2}=2$ }!
\bigskip
\bigskip
\bigskip
\bigskip

\centerline {\epsfxsize 3.5 in \epsfysize 2.5 in\epsfbox{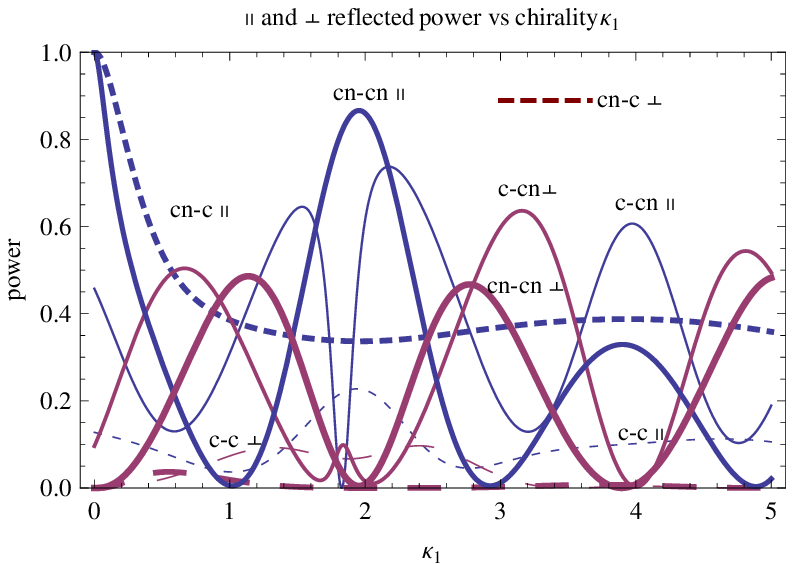}}
 \vskip -0.1 in
\caption Figure 2.6.! {Behavior of reflected power for parallel and
perpendicular field components. $\kappa_{2}=.25,$ $\theta_{i}=\pi/4$
and $\eta_{1} $=$\eta_{2}=2$ }!
 \vfill\eject

Figure~2.6 shows plots of power for impedance matching case versus
chirality of the first layer.  For the purpose of analysis chirality
values are considered between 0 and 5. For cn-c case, almost total
power reflection in terms of parallel component occurs at very low
values of chirality and perpendicular component is almost zero for
considered range of chirality. For c-cn case, zero reflection of
parallel power component is obatined at chirality value
$\kappa_1=1.9$ whereas at $\kappa_1=1.8, 2.2, 4$ perpendicular
component of reflected power is zero. In cn-cn case, oscillating
behavior is observed for both components with periodic behavior for
cn-cn$_{\bot}$. At $\kappa_1= 0, 2, 4$ perpendicular component of
power becomes zero and at $\kappa_1=1, 3, 5$ parallel component of
power is zero. At $\kappa_1=0$, total reflection occurs. Zeros of
parallel and perpendicular components of power at specific values of
chirality can be used to get or avoid particular component of
polarization.

Figure 2.7 gives transmission behavior of power. It is noted that
both components for each case contain either zeros or values
approaching to zero. These findings may be used get or avoid
particular component of polarization. In Figure~2.8, no Brewster
angle exist for all four cases except for cn-c case where both
components are negligibly small at $k_2=1.9$. It is also noted that
perpendicular components of power for cn-cn and cn-c are negligibly
small. Conclusions similar to Figure~2.7 may be drawn from
Figure~2.9.

\centerline{}\centerline{}\centerline {\epsfxsize 3.5 in \epsfysize
2.5 in\epsfbox{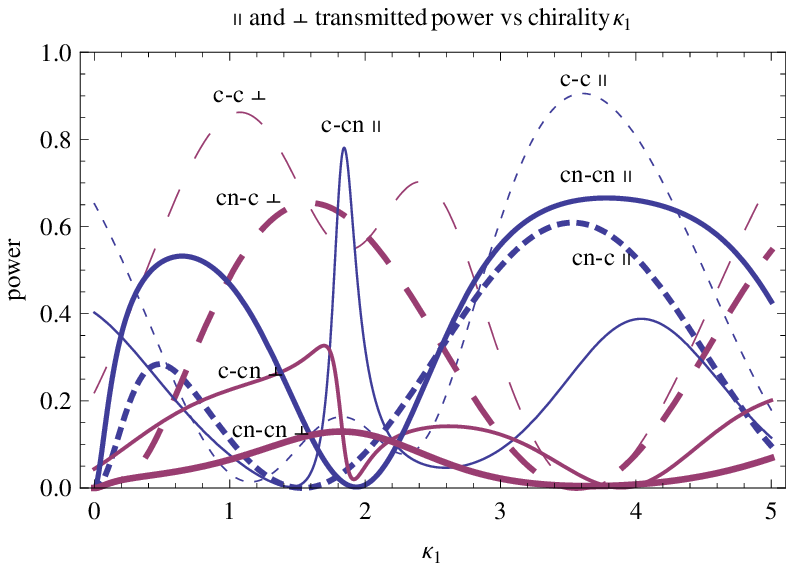}}
 \vskip -0.1 in
\caption Figure 2.7.! {Behavior of transmitted power for parallel
and perpendicular field components. $\kappa_{2}=.25,$
$\theta_{i}=\pi/4,$ $\eta_{1} $=$\eta_{2}=2$ }!

\bigskip

Figure 2.10 shows the plots for reflected power assuming very high
value of permittivity of second layer. This situation may be
considered as a chiral layer backed by perfect electric conductor
interface. In general reflected power has both components, i.e.,
both ${\rm c-PEC}_{\|}$ and ${\rm c-PEC}_{\bot}$ are nonzero. But
for certain values of chirality of first layer, reflected field
contains only co-polarized component. Contribution of parallel
component is very large as compared to the perpendicular component
of reflected power. For cn-PEC case, complete reflection in terms of
parallel component is observed. Figure~2.11 shows zero transmission
of power for ${\rm c-PEC}$ and ${\rm cn-PEC}$ cases.

Another interesting case  is shown in Figure 2.12 where second layer
is assumed of nihility metamaterial. This has been achieved by
taking $\kappa_2=0$ and setting very low value of permittivity of
second layer. This situation may be seen as a chiral layer backed by
nihility interface. For c-n case, almost total reflection in terms
of parallel  component of the reflected power is observed. For the
case of cn-n, perpendicular component of reflected power has also
significant contribution for certain range of chirality. It is noted
that for certain range of chirality, overlap of cn-n and c-n cases
happen. Figure 2.13 shows plots of transmitted power for c-n and
cn-n cases and no transmission of power is obtained. It may be noted
that conclusions drawn from Figure~2.10 to Figure~2.13 agree with
published work.
 \vfill\eject
\centerline {\epsfxsize 3.5 in \epsfysize 2.5 in\epsfbox{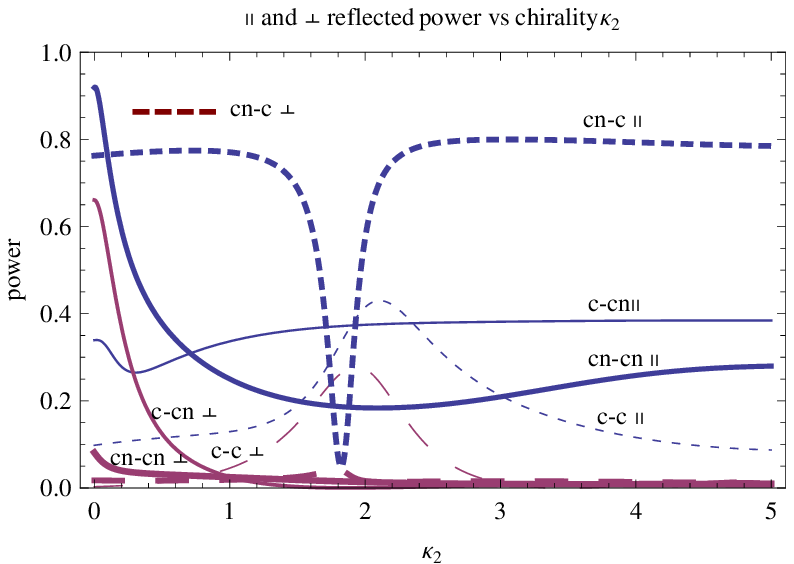}}
 \vskip -0.1 in
\caption Figure 2.8.! {Behavior of reflected power for parallel and
perpendicular field components. $\kappa_{1}=.25,$
$\theta_{i}=\pi/4,$ $\eta_{1}=\eta_{2}=2$ }!

\bigskip
\bigskip
\bigskip
\bigskip
\centerline {}\centerline {} \centerline {\epsfxsize 3.5 in
\epsfysize 2.5 in\epsfbox{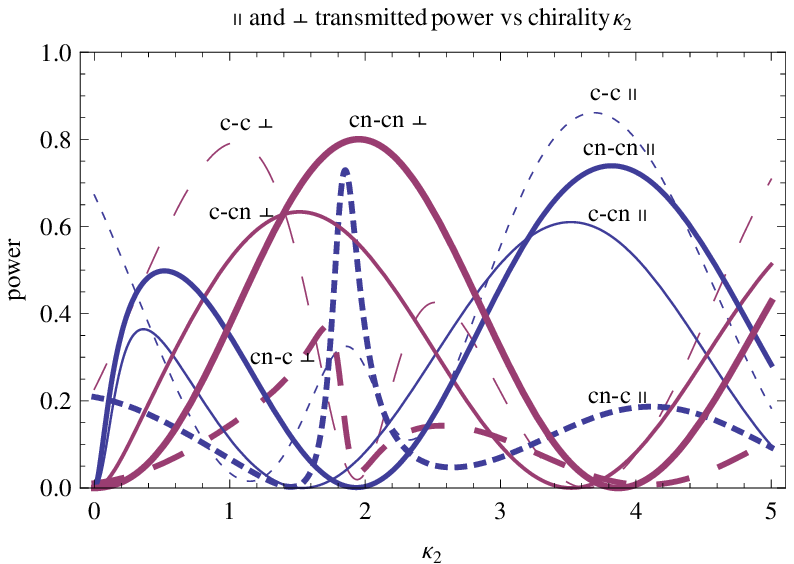}}
 \vskip -0.1 in
\caption Figure 2.9.! {Behavior of transmitted power for parallel
and perpendicular field components. $\kappa_{1}=.25$,
$\eta_{1}$=$\eta_{2}=2$,  and $\theta_{i}=\pi/4$    }!

 \vfill\eject \centerline {\epsfxsize
3.5 in \epsfysize 2.5 in\epsfbox{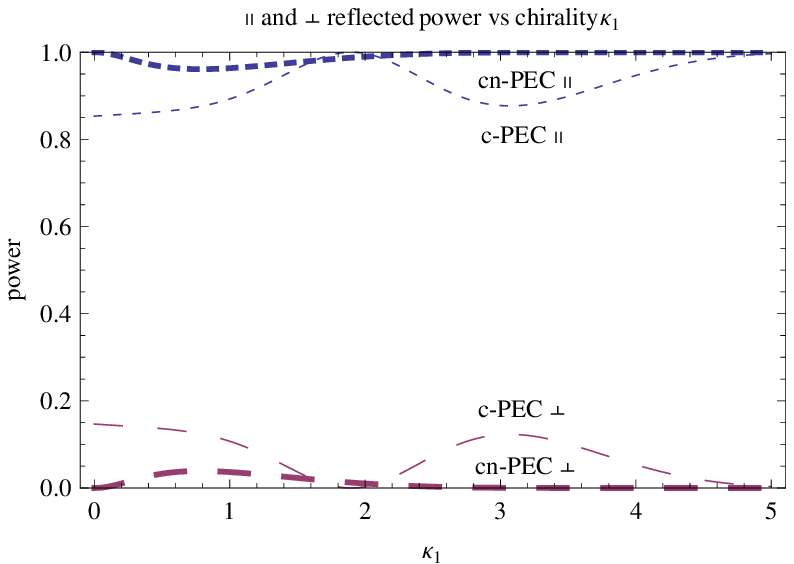}}
 \vskip -0.1 in
\caption Figure 2.10.! {Behavior of reflected power for parallel and
perpendicular field components. $\theta_{i}=\pi/4,$ $\eta_{1} =2$,
$\epsilon_2=10^5$, $\kappa_2=0$ }!

\bigskip
\bigskip
\bigskip
\bigskip
\centerline {}\centerline {} \centerline {\epsfxsize 3.5 in
\epsfysize 2.5 in\epsfbox{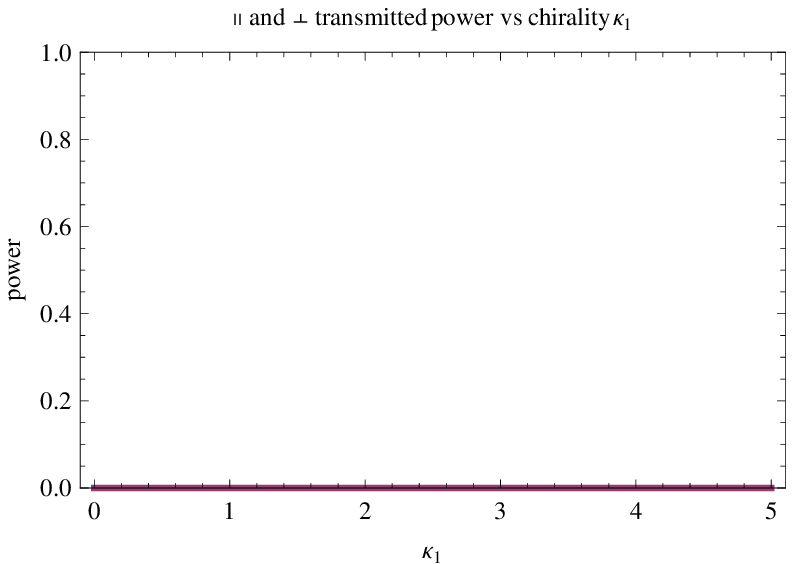}}
 \vskip -0.1 in
\caption Figure 2.11.! {Behavior of transmitted power for parallel
and perpendicular field components. $\theta_{i}=\pi/4$, $\eta_{1}
=2$, $\epsilon_2=10^{5}$, $\kappa_2=0$  }!

\vfill\eject \centerline {\epsfxsize 3.5 in \epsfysize 2.5
in\epsfbox{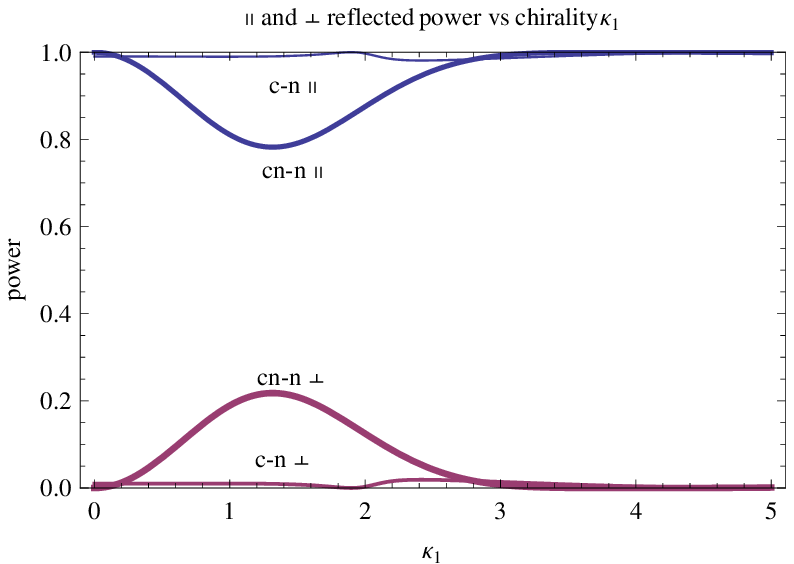}}
 \vskip -0.1 in
\caption Figure 2.12.! {Behavior of reflected power for parallel and
perpendicular field components. $\theta_{i}=\pi/4$, $\eta_{1}=2 $,
$\epsilon_2=10^{-5}$, $\kappa_2=0$   }!

\bigskip
\bigskip
\bigskip
\bigskip
\centerline {}\centerline {} \centerline {\epsfxsize 3.5 in
\epsfysize 2.5 in\epsfbox{Fig11.eps}}
 \vskip -0.1 in
\caption Figure 2.13.! {Behavior of transmitted power for parallel
and perpendicular field components. $\theta_{i}=\pi/4$, $\eta_{1}
=2$, $\epsilon_2= 10^{-5}$, $\kappa_2=0$  }!
 \vfill\eject

Focusing of electromagnetic waves from a paraboloidal reflector
composed of chiral and/or chiral nihility metamaterial using the
Maslov's method is studied in next section. PEC and nihility backed
chiral/chiral nihility paraboloidal reflector are also discussed.
\section{3.0 Evaluation of Finite Field Around Focus}
Considered a bilayer paraboloidal reflector placed in air as shown
in Figure~3.0. The equation for the surface of a paraboloidal
reflector is given by,
$$ \zeta=f(\xi,\eta)=f-{\rho^{2}\over
4f}=f-{\xi^{2}+\eta^{2}\over
 4f}\eqno(26)$$ where $(\xi,\eta,\zeta)$ are the Cartesian coordinate of the point
on the surface of paraboloidal reflector. $f$ is the focal length of
the paraboloidal reflector and $\rho^{2}= \xi^{2}+\eta^{2}$.

The incident field traveling along $z$ axis is expressed as,
$${\bf{E}}_i={\bf{a}}_x \exp{(-ik_0z)} \eqno(27)$$  The incident
plane wave make an angle $\alpha$ with surface normal, where surface
normal is given as, $$\eqalignno
{{\bf{a}}_{n}=\sin{\alpha}\cos{\gamma}{\bf{a}}_{x}
+\sin{\alpha}\sin{\gamma}{\bf{a}}_{y}+\cos{\alpha}{\bf{a}}_{z} &&
(28)}$$ where $\alpha$ and $\gamma$ are given as,
$$\eqalignno{\sin{\alpha}&={\rho\over \sqrt{\rho^{2}+4f^{2}}}, &(29)\cr \cos\alpha &=
{{2f}\over {\sqrt{\rho^{2}+4f^{2}}}}, & (30)\cr \tan{\gamma} &=
{\eta\over\xi}& (31)}$$

\vskip -1 in \centerline {\epsfxsize 8.5 in \epsfysize 6.0
             in\epsfbox{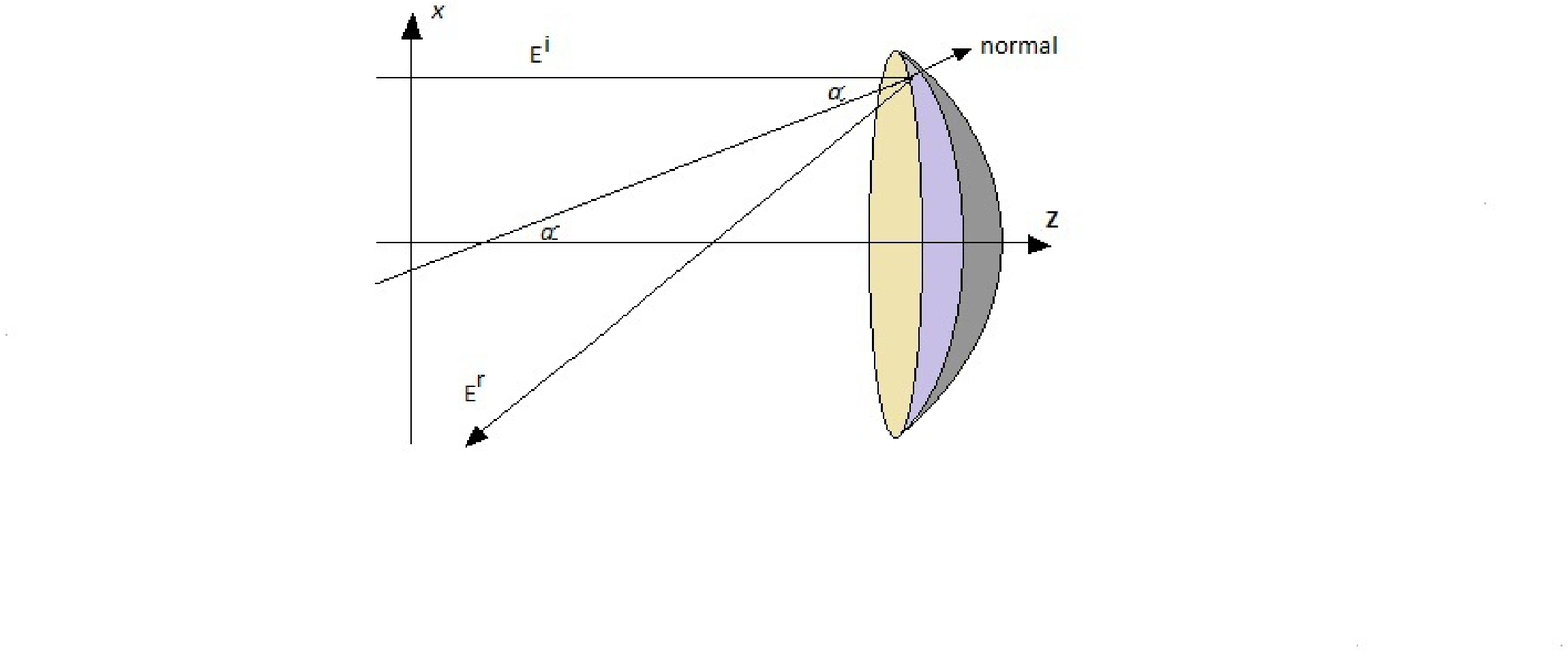}}
             \vskip -1.5 in \caption Figure 3.0.! {Reflection from a paraboloidal
             reflector placed in air
              }!

\noindent The wave reflected from paraboloidal reflector is given
below,
$$ {\bf E}_r= {\bf E}(r_0) \exp[ik_0(x\sin
2\alpha\cos\gamma+y\sin2\alpha\sin\gamma+z\cos 2\alpha)] \eqno(32)$$
Rectangular  components of field ${\bf E}(r_0)$, at the surface of
reflector, are given below,
$$ \eqalignno {E_{x0} &
=B_\perp\sin^2\gamma-B_\|\cos^2\gamma\cos2\alpha & (33)\cr E_{y0} &
=-\cos\gamma\sin\gamma(B_\|\cos2\alpha+B_\perp) &(34)\cr E_{z0}&
=B_\|\sin2\alpha\cos\gamma &(35)}$$ where $B_\|$ and $B_\bot$ are
parallel and perpendicular components of GO reflected field from
planar layers of chiral and/or chiral nihility layers. The field
expression in polar coordinates and valid around the focal point is
given below~[40],
$$ \eqalignno{&{\bf E}_r
={i2k_0f\over \pi}\int_0^H\int_0^{2\pi}\cr &\times{\bf E}_{r0}
\tan\alpha
\exp[-ik_0(2f-r\sin\theta\sin2\alpha\cos(\phi-\gamma)-r\cos\theta\cos2\alpha)]d
\alpha d \gamma  &(36) }$$ The upper limit of integration with
respect to $\alpha$ is taken as,
$$\eqalignno {  H=\tan^{-1}{\left({D\over 2f}\right)}&& }$$
where $D$ is the height of the paraboloidal reflector from
horizontal axis.
\section{3.1 Numerical Results and Discussion}
The field behavior around the focal region of a paraboloidal
reflector are obtained by solving the equation, obtained using
Maslov's method. Numerical results are presented with the variation
of different parameters, i.e., chirality, relative permittivity and
thickness of the layer. Chiral nihility is introduced as limiting
case of chiral medium where value of relative permittivity and
relative permeability of the chiral medium is assumed as $10^{-5},$
keeping nonzero chirality whereas for nihility material, the value
of chirality parameter is taken equal to zero. In all the
simulations, parameters ${(k_{0}, f, H)}$ are taken as ${(1, 100,
\pi/4)}$ . For all cases, permeability of each chiral layer is taken
as unity. All plots deal with behavior of $|{\bf U}(r)|$ versus
$kz$.
\subsection{3.1.1 Chiral-Chiral Paraboloidal Reflector}
 From Figure~3.2 and Figure~3.3, it is noted that by increasing the
permittivity of each layer of the paraboloidal reflector, increases
the strength of electric field around the focus.
 Comparison of Figures 3.2 and 3.3, also shows that strength of field around focus is
 more dependent on the permittivity of second layer.

 In Figure~3.4 and Figure 3.5 , effects of variation of thickness of
 each layer of paraboloidal reflector are presented. It is noted that
increase of thickness of first or second layer also increases the
strength of field around focus. It is noticed that first layer has
more impact on the field strength as compared to the variation of
thickness of the second layer.

 Figures~3.6 and 3.7 show the
effect of change of chirality parameter of each layer while keeping
all the other parameters fix. It is observed that increase in
chirality of each layer also strengthen the field around focal point
of the paraboloidal reflector. Change in chirality of first layer
produces greater change in reflected field strength around the focal
point as compared to chirality variation for second layer.

For the case of  c-c paraboloidal reflector, some interesting
features have been studied that are permittivity, chirality and
thickness of each chiral layer is directly proportional to the
reflector's field strength at focal point. Moreover it has been
observed that, first and second layers show different effects
against the variation of a particular parameter while keeping all
the other parameters fix.
\section{3.1.2 Chiral-PEC and Chiral-Nihility  Paraboloidal Reflector }
In this section two special cases  for paraboloidal reflector are
discussed one  by one: paraboloidal reflector composed of c-PEC and
c-n cases. Figure~3.8 shows reflected field strength for three
different values of permittivity of the first layer when second
layer is considered as PEC (very large value of permittivity, i.e.,
$10^5$). It is noted that increase in permittivity of first layer
also increases the field strength at the focal point. Figure~3.9
shows that increase in the chirality of first layer correspondingly
increases the field strength at the focus. In Figures 3.10 and 3.11,
second layer is considered as nihility (by assuming very small
values of permittivity and permeability, i.e., $10^{-5}$). From
Figures 3.10 and 3.11, it is observed that by increasing the values
of permittivity and chirality, the field strength around the focal
point also increases correspondingly.

\section{4.0 Conclusions}
Characteristics  of reflected and transmitted  powers from a
structure of two parallel layers filled with chiral and/or
chiral-nihility metamaterials are studied. Special cases of PEC and
nihility backed chiral/chiral nihility layer are also discussed. It
is observed that the behavior of reflected or transmitted powers
strongly depend on the angle of incidence and values of constitutive
parameters describing the matamaterials. For some specific values
and  ranges of incident angles, total reflection of power has been
observed hence structure can yield range of Brewster angles.
Parallel or perpendicular component of reflected/transmitted power
may be selected. That is, either co-polarized or cross polarized
field is produced. Similarly for some values of chirality, total
reflection/ transmission of power is also observed. This arrangement
of parallel layers can also be used as a power divider.

Arrangement of two parallel layers mentioned above is utilized to
study the behavior of field around the focal region of a large size
paraboloidal reflector. Paraboloidal reflector, composed of c-c,
c-PEC and c-n layers is discussed for this purpose. PEC and nihility
backed cases give higher values of reflected field around the focus
as compared to composed of chiral-chiral layers. It is concluded
that increase in the relative permittivity of each layer also
increases field strength around the focus. An increase in the
thickness and chirality parameter of the layer also have the same
effect, i.e., increase in values of these parameters yields
correspondingly higher focusing of the field. It is also noted that
different parameters, mentioned above, affect the strength of the
field differently. Moreover in the case of c-c paraboloidal
reflector, variation of a parameter for first and second layer
yields different reflected field strength when all other parameters
are kept constant. In this case, change in permittivity of second
layer produces more variation in the reflected field strength around
the focal point as compared to the first layer, whereas for change
in chirality and thickness, field strength of the reflector is more
sensitive for the first layer. Hence from above discussion strength
of the reflected field around the paraboloidal reflector can be
controlled  by studying the effect of different factors. Study of
focusing  field strength is of great importance in defence and other
electronic systems. This study provides an investigation to design
an efficient reflector coated with materials having different
material properties. This whole discussion suggests to get control
over the focusing of the reflected field  by bringing variations in
the permittivity, thickness or the chirality of the corresponding
layer.

\vfill\eject
 \vskip .1 in \hskip -.5 in\centerline
{\epsfxsize 3.5 in \epsfysize 2.5 in\epsfbox{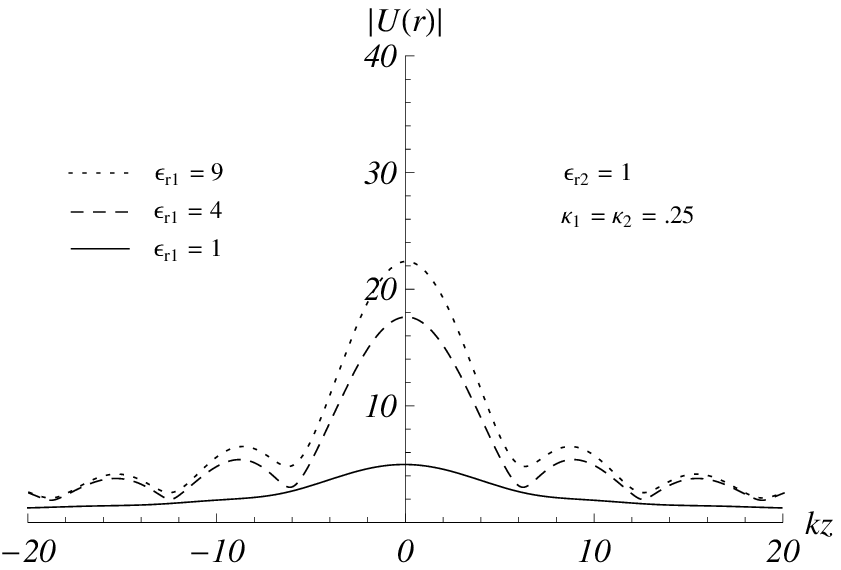}}
 \vskip -0.1 in\caption Figure 3.2.! { Field behavior around focus for different values of permittivity
of first layer:  $d_1=d_2={\lambda_0\over 4},$ {\rm c-c case.} }!

\vskip .1 in \hskip -.5 in  \centerline {\epsfxsize 3.5 in
\epsfysize 2.5 in\epsfbox{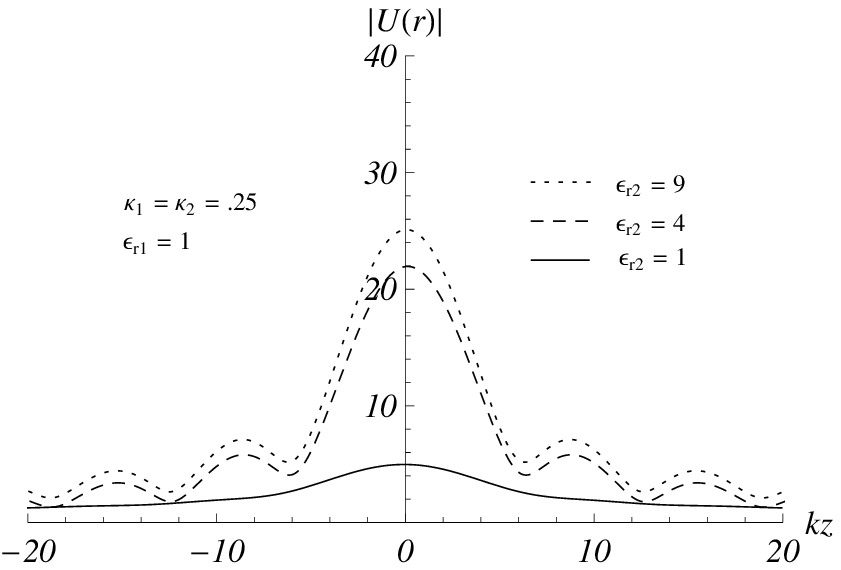}} \vskip -0.1 in \caption Figure
3.3.! { Field behavior around focus for different values of
permittivity of second layer:  $d_1=d_2={\lambda_0\over 4},$ {\rm
c-c case.} }!

\vfill\eject \hskip -.5 in \centerline {\epsfxsize 3.5 in \epsfysize
2.5 in\epsfbox{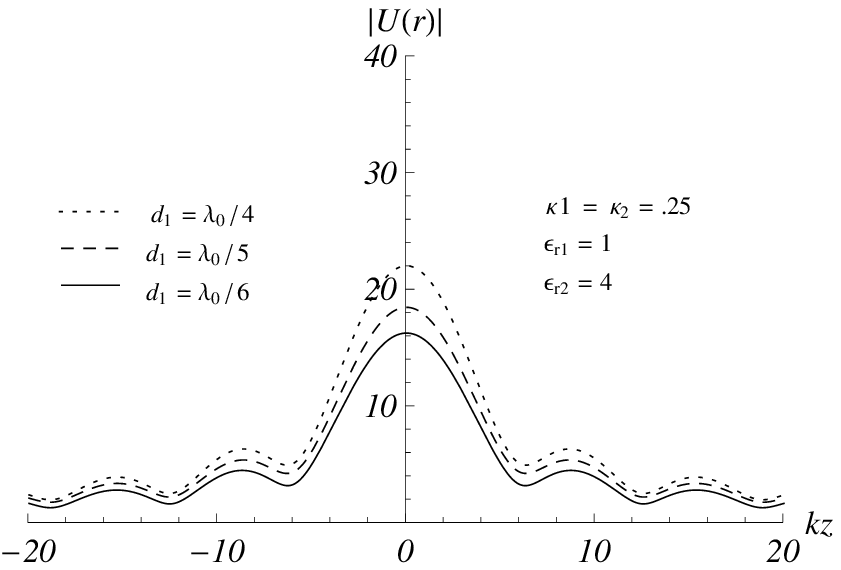}} \vskip -0.1 in \caption Figure 3.4.! {
Field behavior around focus for different values of thickness of
first layer: $d_2={\lambda_0\over 4},$ {\rm c-c case.} }!
\bigskip
 \hskip -.5 in \centerline {\epsfxsize 3.5 in \epsfysize 2.5
in\epsfbox{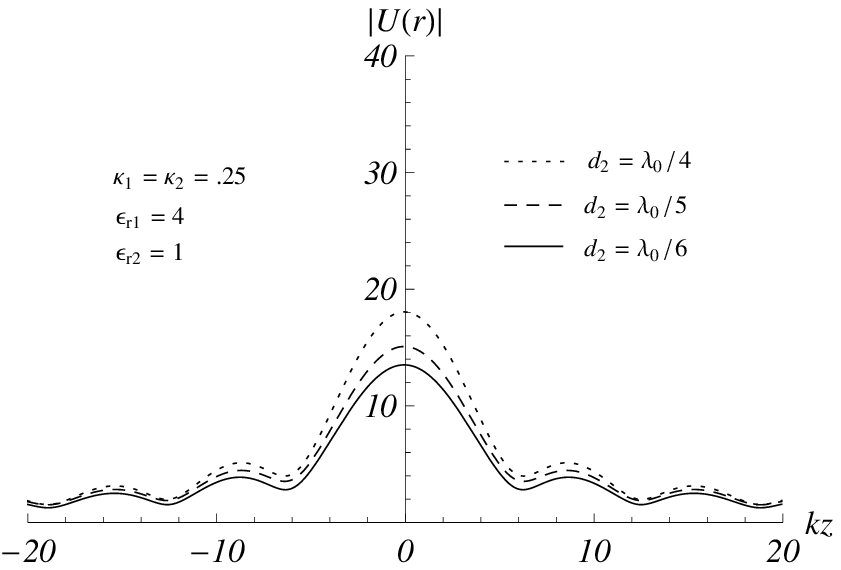}}
 \vskip -0.1 in
\caption Figure 3.5.! {Field behavior around focus for different
values of thickness of second layer: $d_1={\lambda_0\over 4}$, {\rm
c-c case.} }!

\vfill\eject
 \centerline
{\epsfxsize 3.5 in \epsfysize 2.5 in\epsfbox{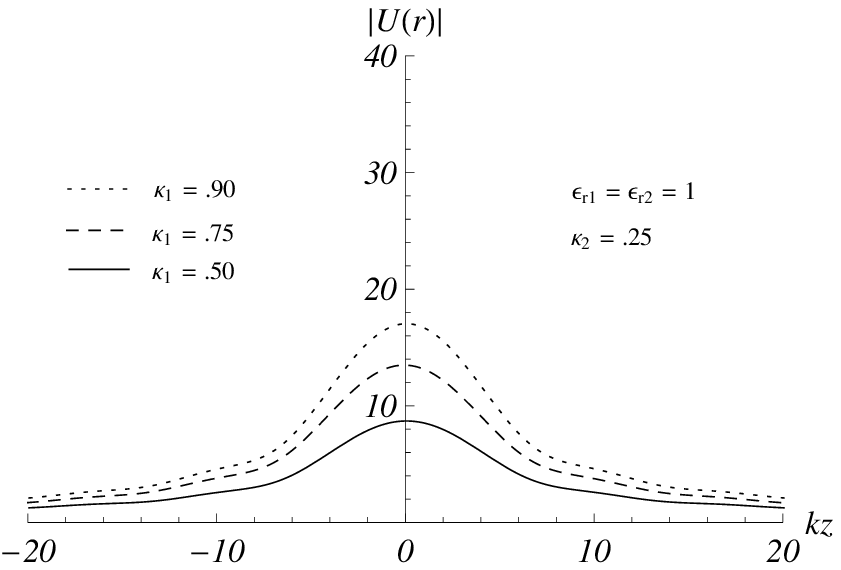}}
 \vskip -0.1 in
\caption Figure 3.6.! {Field behavior around focus for different
values of chirality of first layer: $d_1=d_2={\lambda_0\over 4}$,
{\rm c-c case.}}!
\bigskip

\centerline {\epsfxsize 3.5 in \epsfysize 2.5 in\epsfbox{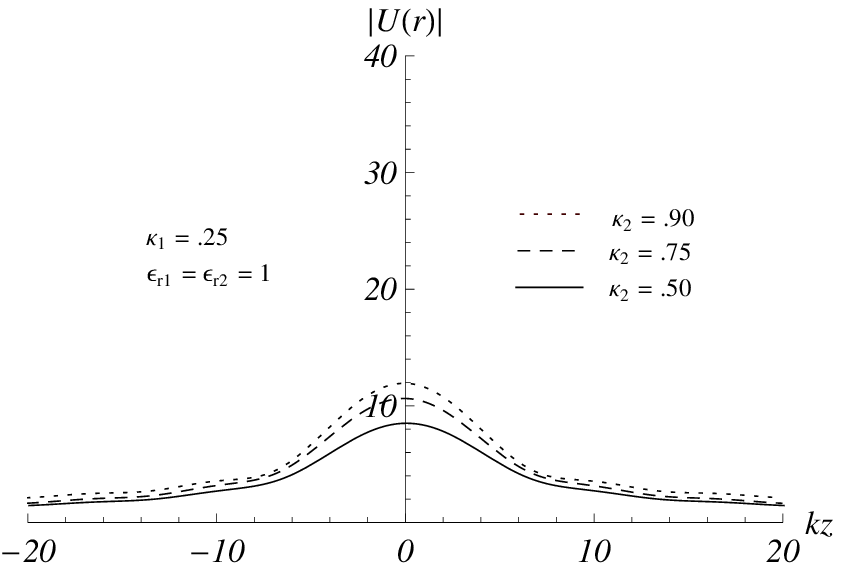}}
 \vskip -0.1 in
\caption Figure 3.7.! {Field behavior around focus for different
values of chirality parameter of second layer:
$d_1=d_2={\lambda_0\over 4}$,
 {\rm c-c  case.} }!
\bigskip

\vfill\eject \vskip .1 in
 \centerline {\epsfxsize 3.5 in \epsfysize 2.5 in
\epsfbox{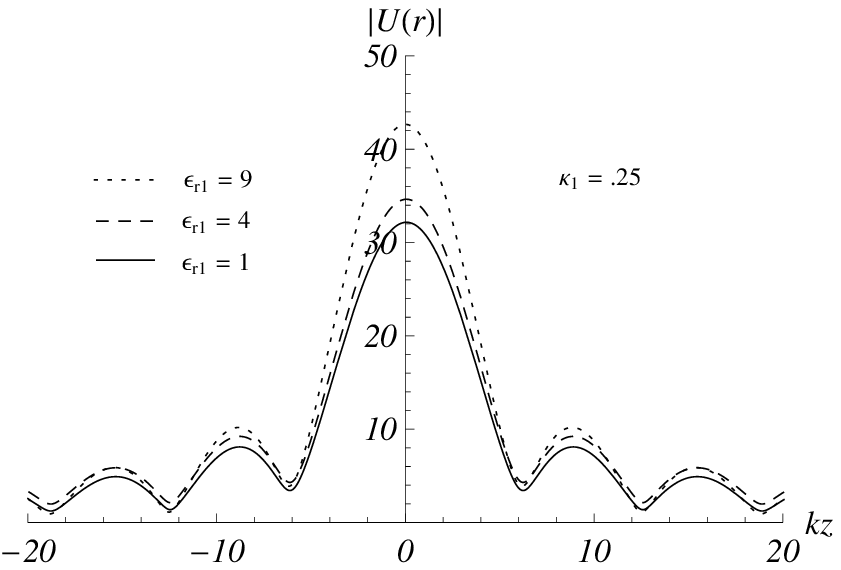}}
 \vskip -0.1 in
 \caption Figure 3.8.! { Field behavior around focus for different values of permittivity
of first layer: $d_1=d_2={\lambda_0\over 4}$, $\epsilon_{r2}=1\times
10^5,$ {\rm c-PEC case.} }!

\vskip .1 in
 \centerline {\epsfxsize 3.5 in \epsfysize 2.5 in
\epsfbox{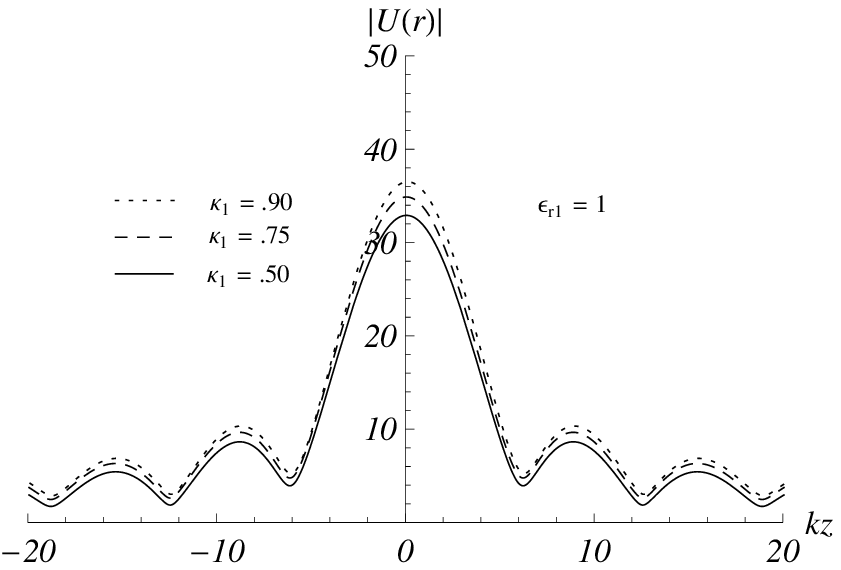}}
 \vskip -0.1 in
 \caption Figure 3.9.! { Field behavior around focus for different
 values of chirality of first layer: $d_1=d_2={\lambda_0\over 4}$,
   $\epsilon_{r2}=1\times 10^5,$ {\rm c-PEC
case.}
              }!
\vfill\eject

 \vskip .1 in
 \centerline {\epsfxsize 3.5 in \epsfysize 2.5 in
\epsfbox{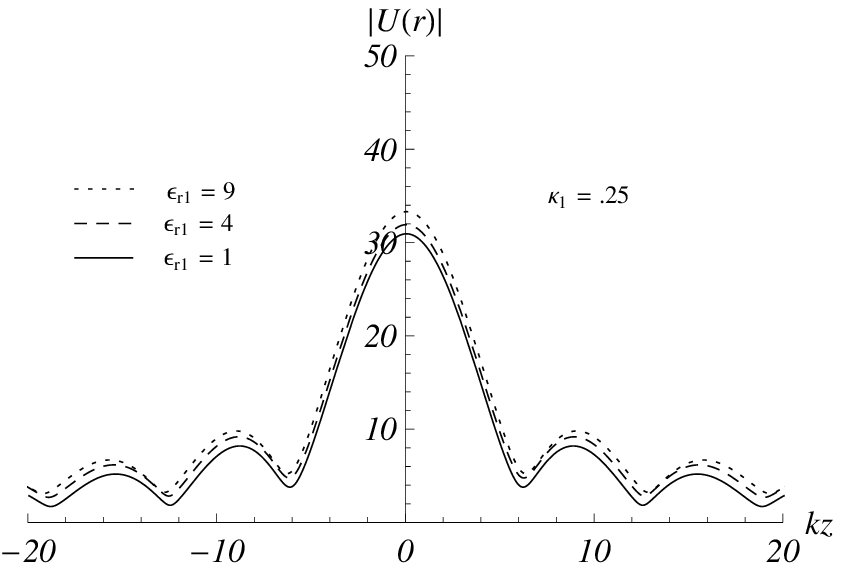}}
 \vskip -0.1 in
 \caption Figure 3.10.! {Field behavior around focus for different values of permittivity
of first layer: $d_1={\lambda_0\over 4}$, $\epsilon_{r2}=1\times
10^{-5},$ $\kappa_2=0$, {\rm c-n case.}
              }!

\vskip .1 in
 \centerline {\epsfxsize 3.5 in \epsfysize 2.5 in
\epsfbox{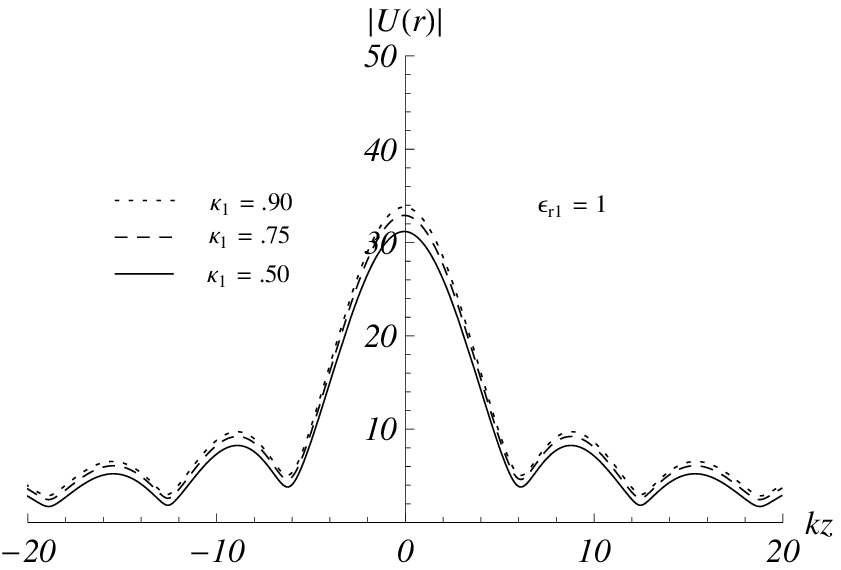}}
 \vskip -0.1 in
 \caption Figure 3.11.! { Field behavior around focus for different values of permittivity
of first layer: $d_1={\lambda_0\over 4}$, $\epsilon_{r2}=1\times
10^{-5},$ $\kappa_2=0$, {\rm c-n case.} }! \vfill\eject

\vfill\eject

{ \bigbold References}\vskip -0.1 in
\bigskip

\item{1.} D. F. Arago, ``Sur une modification remarquable qu' eprouvent les
rayons lumineux dans leur passage a travers certains corps
diaphanes, et sur quelques autres nouveaux phenomnnes d'optique",
Mem. Inst. 1, 93-134 (1811).
\item{2.} J. B. Biot, ``Ph6nomenes de polarisation
successive, observs dans des fluides homogenes", Bull. Soc.
Philomath., 190-192 (1815).
\item{3.} A. Fresnel, ``Memoire sur la double refraction que les rayons
lumineux eprouvent en traversant les aiguilles de cristal de roche
suivant des directions parall6les a l'axe", Oeuvres 1, 731- 751
(1822).
\item{4.} J. B. Biot, ``M6moire sur la polarisation circulaire et sur ses
applications a la chimie organique", Mem. Acad. Sci. 13, 39-175
(1835).
\item{5.}K. F. Lindman, ``Ober eine durch ein isotropes system von
spiralformigen resonatoren erzeugte rotationspolarisation der
elektromagnetischen wellen", Ann. Phys. 63, 621-644 (1920).
 \item{6.} K. F. Lindman, ``Uber die durch- ein aktives raumgitter erzeugte
rotationspolarisation der elektromagnetischen wellen", Ann. Phys.
69, 270-284 (1922).
\item{7.} D. L. Jaggard, A. R. Mickelson, C. H. Papas, ``On electromagnetic
 waves in chiral media",  Appl. Phys. 18, 211-216 (1979).
\item{8.} I. V. Lindell, A. H. Sihvola, S. A. Tretyakov,
   A. J. Viitanen,
   ``Electromagnetic waves in chiral and bi-isotropic media",
   Artech
   House, Boston (1994).
\item{9.} S. Bassiri, C. H. Papas, and N. Engheta , ``Electromagnetic wave
propagation through a dielectric-chiral interface and through a
chiral slab",  J. Opt. Soc. of Am. A,  5, 1450-1459 (1988).
\item{10.} A. Lakhtakia, ``Beltrami fields in chiral media", World Scientific, Singapore(1994).
\item{11.} V. G. Vaselago, ``The electrodynamics of substances with
   simultaneously negative values of permittivity and permeability",
   Sov. Phys. Usp. 10,  509-514 (1968).
\item{12.} C. W. Qiu, H. Y. Yao, S. Zouhdi, L. W. Li, and M. S. Leong,
``On the constitutive relations of G-chiral media and the
possibility to realize negative-index media", Microwave Opt.
Technol. Lett. 48, 2534--2538 (2006).
\item{13.} S. Tretyakov, A. Sihvola, and L. Jylhä, ``Backward-wave regime and
negative refraction in chiral composites", Photonics Nanostruct.
Fundam. Appl. 3, 107--115 (2005).
\item{14.} J. B. Pendry, A. J. Holden, W. J. Stewart, and I. Youngs,
``Extremely low frequency plasmons in metallic mesostructures" Phys.
Rev. Lett. 76, 4773–-4776 (1996).
\item{15.} J. B. Pendry, ``A chiral route to negative refraction" Science
306, 1353–-1355 (2004).
\item{16.} C. W. Qiu, H. Y. Yao, L. W. Li, S. Zouhdi, and T. S. Yeo,
``Routes to left-handed materials by magnetoelectric couplings"
Phys. Rev. B 75, 245214 (2007).
\item{17.} S. Zhang, Y. S. Park, J. Li, X. Lu, W.
Zhang, and X. Zhang,``Negative refractive index in chiral
metamaterials", Phys. Rev. Lett. 102, 023901 (2009).
\item{18.} T. G. Mackay, A. Lakhtakia, Simultaneous negative- and positive-phase-velocity
propagation in an isotropic chiral medium, Microwave and Optical
Technology Letters 49, 1245–-1246 (2007).
\item{19.} A. Lakhtakia,  ``An electromagnetic trinity from negative permittivity
and negative permeability", Int. J. Inf. and Mil. Wav., 22,
1731-1734 (2001)
\item{20.} S. Tretyakov, I. Nefedov, A. H. Sihvola, S. Maslovki, C.
Simovski,`` Waves and energy in chiral nihility", Journal of
Electromagnetic Waves and Applications 17, 695-706 (2003).
\item{21.} Q. A. Naqvi,
``Planar slab of chiral nihility metamaterial backed by fractional
dual/pemc interface", Progress In Electromagnetics Research PIER 85,
381–-391 (2008).
\item{22.} M. A. Baqir, A. A. Syed and Q. A. Naqvi,
"Electromagnetic fields in a circular waveguide containing chiral
nihility metamaterial" Progress In Electromagnetics Research M 16,
85--93 (2011).
\item{23.} C. A. Balanis, ``Advanced engineering electromagnetics",
John Willey and Sons, 2nd Edition (2012).
\item{24.} D. K. Cheng, ``Fields and wave electromagnetics", Addison-Wesley, Newyork (1989).
\item{25.} C. W. Qiu, N. Burokur, S. Zouhdi,
and L. W. Li, ``Chiral nihility effects on energy flow in chiral
materials", J. Opt. Soc. of Am.,  25, 55-63 (2008).
\item{26.} F. Ahmad, S. N. Ali, A. A. Syed, and Q. A.
Naqvi,``Chiral and/or chiral nihility interfaces: parametric
dependence, power tunneling and rejection" Progress in
Electromagnetics Research M  23, 167--180 (2012).
\item{27.} L. B. Felson,  Hybrid formulation of wave Propagation and
scattering, Nato ASI Series, Martinus Nijho, Dordrecht, The
Netherlands, (1984).
\item{28.} G. A. Dechamps,  ``Ray techniques in electromagnetics," Proc.
IEEE, 60, 1022-1035 (1972).
\item{29.} C. H. Chapman  and R. Drummond, ``Body wave seismograms in
inhomogeneous media using Maslov asymptotic theory," Bull. Seismol.,
Soc. Am., 72, 277-317, (1982).
\item{30.} V. P. Maslov,  ``Perturbation theory and asymptotic method" ,
Gos. Moskov.,  Univ., Moscow, (1965) (in Russian). (Translated into
Japanese by Ouchi et al., Iwanami, Tokyo, 1976).
\item{31.} A. Ghaffar, Q. A. Naqvi, and K. Hongo, ``Analysis of the fields
in three dimensional Cassegrain system", Progress In
Electromagnetics Research PIER 72, 215–-240, (2007).
\item{32.} Y. Ji,  and K. Hongo, ``Analysis of
electromagnetic waves refracted by a spherical dielectric interface
by Maslov's method", J. Opt. Soc. Am. A, 8, 541–-548 ( 1991).
\item{33.}
Y. Ji, and K. Hongo, ``Field in the focal region of a dielectric
spherical by Maslov's method", J. Opt. Soc. Am. A,  8, 1721–-1728
(1991).
\item{34.} K.Hongo, , Y. Ji, and E. Nakajima, ``High frequency
expression for the field in the caustic region of a reflector using
Maslov's method", Radio Sci. 21, 911–-919 (1986).
\item{35.}
K. Hongo,  and Y. Ji, "High frequency expression for the field in
the caustic region of a cylindrical reflector using Maslov's
method", Radio Sci. 22, 357–-366 (1987).
\item{36.} K. Hongo,  and Y.
Ji, ``Study of the field around the focal region of spherical
reflector antenna by Maslov's method", IEEE Trans. Antennas
Propagat.,  36, 592–-598 (1988).
\item{37.} R. W. Ziolkowski,  and G.
A. Deschamps, ``Asymptotic evaluation of high frequency field near a
caustic: an introduction to Maslov's method", Radio Sci. 19,
1001–-1025 (1984).
\item{38.} M. Faryad,  and Q. A. Naqvi, ``High frequency
expression for the field in the caustic region of cylindrical
reflector placed in chiral medium", Progress In Electromagnetics
Research PIER 76, 153–-182 (2007).
\item{39.} M. Faryad,  and Q.
A. Naqvi, ``High frequency expression for the field in the caustic
region of a parabolic reflector coated with isotropic chiral
medium," Journal of Electromagnetic Waves and Applications 22,
965–-986 (2008).
\item{40.} T. Rahim, , M. J. Mughal, Q. A. Naqvi, and M. Faryad, ``Focal region field of a
paraboloidal reflector coated with isotropic chiral medium" Progress
In Electromagnetics Research PIER 94, 351-366 (2009).
\item{41.} A. Illahi and Q. A. Naqvi ``Study of focusing of electromagnetic waves reflected by a PEMC
backed chiral nihility reflector using Maslov's method", Journal of
Electromagnetic Waves and Applications  23, 863-873 (2009).
\item{42.} C. Sabah, S. Uckun ``Mirrors with chiral slabs"
Journal of Optoelectronics and Advanced Materials 8, 1918-1924
(2006).

\vfill\eject

\bye